\documentclass[
superscriptaddress,
twocolumn,
 amsmath,
 amssymb,
 prl,
]{revtex4-2}
\usepackage{import}
\usepackage{lipsum}
\usepackage{braket}
\usepackage{graphicx}
\usepackage{dcolumn}
\usepackage{siunitx}
\usepackage[normalem]{ulem}
\usepackage{upgreek}
\usepackage{bm}
\usepackage{color}
\usepackage{notes2bib}
\usepackage{pgf}
\usepackage{natbib}

\definecolor{d_Red}{RGB}{190, 30, 45}
\definecolor{d_Cyan}{RGB}{0, 174, 239}
\definecolor{d_Gold}{RGB}{175, 169, 97}
\definecolor{d_Yellow}{RGB}{255, 213, 58}
\definecolor{d_Purple}{RGB}{104, 36, 109}
\definecolor{d_Heather}{RGB}{203,168,177}
\definecolor{d_Stone}{RGB}{218,205,162}
\definecolor{d_Sky}{RGB}{165,200,208}
\definecolor{d_Cedar}{RGB}{182,170,167}
\definecolor{d_Concrete}{RGB}{179,189,177}
\definecolor{d_Ink}{RGB}{0,42,65}
\definecolor{d_Black}{RGB}{51,49,50 }

\renewcommand{\sout}[1]{}

\begin{document}


\title{A Collectively Encoded Rydberg Qubit}

\author{Nicholas L. R. Spong}%
\affiliation{Joint Quantum Centre (JQC) Durham-Newcastle, Department of Physics, Rochester Building, Durham, DH1 3LE, UK.}

\author{Yuechun Jiao}
\affiliation{Joint Quantum Centre (JQC) Durham-Newcastle, Department of Physics, Rochester Building, Durham, DH1 3LE, UK.}
\affiliation{
 State Key Laboratory of Quantum Optics and Quantum Optics Devices, Institute of Laser Spectroscopy, Shanxi University, Taiyuan 030006, China
}

\author{Oliver D. W. Hughes}%
\affiliation{Joint Quantum Centre (JQC) Durham-Newcastle, Department of Physics, Rochester Building, Durham, DH1 3LE, UK.}

\author{Kevin J. Weatherill}%
\affiliation{Joint Quantum Centre (JQC) Durham-Newcastle, Department of Physics, Rochester Building, Durham, DH1 3LE, UK.}

\author{Igor Lesanovsky}

\affiliation{Institut für Theoretische Physik, Auf der Morgenstelle 14, 72076 Tübingen, Germany}

\affiliation{School of Physics and Astronomy and Centre for the Mathematics and Theoretical Physics of Quantum Non-Equilibrium Systems, The University of Nottingham, Nottingham, NG7 2RD, UK}

\author{Charles S. Adams}
\affiliation{Joint Quantum Centre (JQC) Durham-Newcastle, Department of Physics, Rochester Building, Durham, DH1 3LE, UK.}

\date{\today}

\begin{abstract}
We demonstrate a collectively-encoded qubit based on a single  Rydberg excitation stored in an ensemble of $N$ entangled atoms. Qubit rotations are performed by applying microwave fields that drive excitations between Rydberg states. Coherent read-out is performed by mapping the excitation into a single photon.  Ramsey interferometry is used to probe the coherence of the qubit, and to test the robustness to external perturbations. We show that qubit coherence is preserved even as we lose atoms from the polariton mode, preserving Ramsey fringe visibility. We show that dephasing due to electric field noise scales as the fourth power of field amplitude. These results show that robust quantum information processing can be achieved via collective encoding using Rydberg polaritons, and hence this system could provide an attractive alternative coding strategy for quantum computation and networking. 
\end{abstract}

\maketitle


Quantum technology is increasingly expanding our capabilities in computing, sensing, metrology, and communications. Atomic systems, including those exploiting highly-excited Rydberg states are particularly attractive for quantum applications \cite*{Saffman2010a,Firstenberg2016a,Saffman2016a, Adams2020,Khazali2020,Browaeys2020}, as they offer a unique combination of precision \cite{Bloom2014}, high-fidelity entanglement generation \cite{Jau2016,Levine2018,Levine2019,Graham2019,Madjarov2020}, scaling to 3D \cite{Saffman2008,Barredo2018}, direct photonic read-out \cite{Li2013, Li2019} and strong photon-photon interactions \cite{Pritchard2010a, Dudin2012, Peyronel2012,Maxwell2013,Tiarks2019}. 
Recently, remarkable progress has been made using individual Rydberg atoms for quantum simulation \cite{Bernien2017,Zhang2017, deLeseleuc2019,Scholl2020,Ebadi2020}. In parallel and across the full spectrum of quantum computing platforms, there has been considerable recent interest in the use of collective encoding strategies exploiting different spatial modes \cite{Tordrup2008, Grezes2014, Ranjan2020}, internal states \cite{Brion2007,Brion2008}, grid states \cite{Fluhmann2019,Campagne-Ibarcq2019}, and Schr{\"o}dinger cat states \cite{Grimm2020}.

In this paper, we demonstrate a new collective-coding scheme based on Rydberg polaritons \cite{Lukin2001a, Firstenberg2016a, Adams2020}. 
The novel feature of our scheme is that the qubit is stored as a superposition of Rydberg polariton modes. One advantage of this scheme is that quantum information is distributed over many atoms as opposed to single atom encoding schemes. An additional advantage is that the polariton phase \cite{Fleischhauer2000} enables direct photonic state read out in a well-defined spatial mode \cite{Dudin2012}. Also, the collective character of both qubit states causes the Rabi frequency for qubit rotations to be  independent of the number of atoms. Large transition dipole moments between highly-excited Rydberg states (e.g. the radial matrix element for the $\ket{\rm r} = \ket{60S_{1/2}}$ to  $\ket{\rm r'} = \ket{60P_{3/2}}$ transition is $3684$~Debye \cite{Sibalic2017a}) provide for fast coherent control and SWAP operations \cite{Barredo2015} on time scales of order nanoseconds. Our scheme is scalable to many collective qubits using ensemble arrays  \cite{Wang2020}, and could provide an alternative hybrid strategy for quantum networking exploiting microwave interactions \cite{Kimble2008, VanLeent2019}.

%
\begin{figure}
    \centering
    \includegraphics[page=1]{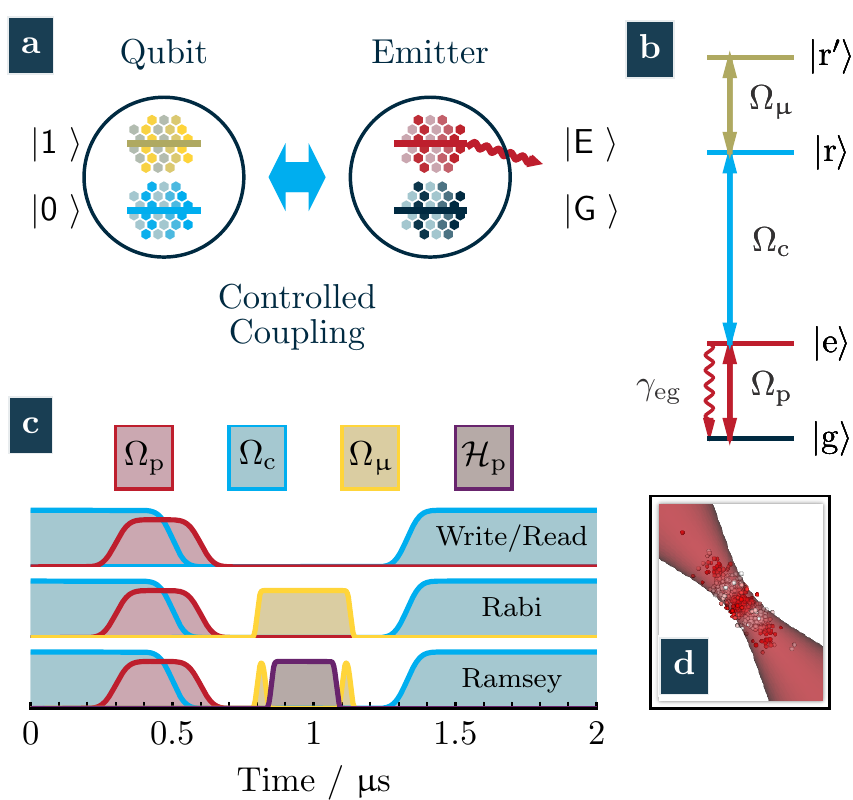}
    \caption{{\bf Collective encoding and read-out}: \textbf{a}: Quantum information is encoded into a Rydberg polariton \cite{Fleischhauer2000} in a superposition of $\vert 0\rangle$ and $\vert 1\rangle$, supported by Rydberg states $\vert {\rm r}\rangle$, $\vert {\rm r}'\rangle$, see Eqs.~(1) and (2).  The qubit coherently couples to a photon emitter, $\vert {\sf E}\rangle,\vert {\sf G}\rangle$, supported by $\ket{\rm e}, \ket{\rm g}$. A control field (blue arrow) provides the coupling with Rabi frequency $\Omega_{\rm c}$. \textbf{b}: The internal states of each atom. Initialisation of the qubit state $\vert {\sf 0}\rangle$ is performed via two-photon excitation of the transition $\vert {\rm g}\rangle\rightarrow\vert {\rm e}\rangle\rightarrow\vert {\rm r}\rangle$. Single excitation of the collective state $\vert 0\rangle$ is enforced by the Rydberg blockade mechanism \cite{Lukin2001a}.  
    Qubit rotations are implemented by driving the transition $\vert {\rm r}\rangle\leftrightarrow\vert {\rm r}'\rangle$ using a microwave field with amplitude characterised by a Rabi frequency $\Omega_\upmu$. Read-out is performed via polariton retrieval from $\vert {\rm r}\rangle\rightarrow\vert {\rm e}\rangle$, whereafter $\vert {\rm e}\rangle$ decays back to the ground state $\vert {\rm g}\rangle$ with rate $\gamma_\mathrm{eg}$ via collective emission into the mode of the original photon.  
    {\bf c}: Pulse sequences used for qubit read/write, Rabi oscillations, and Ramsey interferometry. 
    Perturbative Hamiltonians $\mathcal{H}_{\rm p}$ (purple) can be implemented via external fields. {\bf d}: An illustration of our atomic ensemble in an optical tweezer.}
    \label{fig:schematicAndPulseSequences}
\end{figure}
The main focus of this paper is to demonstrate coherent control of our collective qubit, and to test the robustness of the scheme to both atom loss and decoherence due to environmental noise \cite{Chenu2017}.
The collective encoding scheme works as follows, see Fig.~1: For $N$ atoms within a blockade volume \cite{Lukin2001a}, the transition  $\vert {\rm g}\rangle \rightarrow \vert \rm e \rangle \rightarrow \vert{\rm r}\rangle$ couples the $N$-atom ground state $\vert {\sf G}\rangle =
\vert {\rm g}_1{\rm g}_2\ldots {\rm g}_j \ldots {\rm g}_N\rangle$ to the collective state 
\begin{eqnarray}
\vert {\sf 0}\rangle &=&\frac{1}{\sqrt{N}}\sum_{j=1}^N 
{\rm e}^{
{\rm i}(\boldsymbol{k}\cdot \boldsymbol{R}_j-\omega_{\rm r}t)
}
\vert {\rm g}_1{\rm g}_2\ldots {\rm r}_j \ldots {\rm g}_N\rangle~,
\label{eqn:qubitState0}
\end{eqnarray}
where ${\rm g}_j$ and ${\rm r}_j$ denote atom $j$, with position $R_j$ in states $\vert {\rm g}\rangle$ and $\vert{\rm r}\rangle$, respectively, see Fig.~1. The phase at each atom contains both {\it local phase} terms $ {\rm }\mathbf{k}\cdot\mathbf{R}_j$,  where $\mathbf{k}$ is the effective wave vector of the excitation lasers, and a {\it global phase}, $-\omega_{\rm r}t$, where $\omega_{\rm r}$ is the angular frequency of the transition $\vert{\rm g}\rangle\rightarrow\vert{\rm r}\rangle$.
For an ensemble initialised in $\vert {\sf 0}\rangle$, applying a microwave field with detuning, $\Delta_\mathrm{\upmu}$
relative to the $\ket{\rm r} \rightarrow \ket{{\rm r}'}$ transition, see Fig.~1{\bf a} and {\bf b}, couples $\vert {\sf 0}\rangle$ to the collective state 
\begin{eqnarray}
\vert {\sf 1}\rangle &=&\frac{1}{\sqrt{N}}\sum_{j=1}^N 
{\rm  e}^{{\rm i}
(\boldsymbol{k}\cdot \boldsymbol{R}_j-\omega_{{\rm r}'}t)}
\vert  {\rm g}_1{\rm g}_2\ldots {\rm r}'_j \ldots {\rm g}_N\rangle~.
\end{eqnarray}
As  both $\vert {\sf 0}\rangle$ and $\vert {\sf 1}\rangle$ contain $N$ terms, the Rabi frequency for qubit rotations is independent of the number of atoms, $N$. This enables high-fidelity single-qubit rotations.

Finally, applying a coupling laser (blue in Fig.~1) resonant with the transition 
$\ket{\rm r} \rightarrow \ket{{\rm e}}$ we couple the collective state $\vert {\sf 0}\rangle$ to the state
\begin{eqnarray}
\vert {\sf E}\rangle &=&\frac{1}{\sqrt{N}}\sum_{j=1}^N {\rm  e}^{{\rm i}\boldsymbol{k'}\cdot \boldsymbol{R}_j}\vert  {\rm g}_1{\rm g}_2\ldots {\rm e}_j \ldots {\rm g}_N\rangle~.
\end{eqnarray}
This state decays on a time scale of 10~ns via collective emission of a single photon in a well-defined optical mode \cite{Lukin2001a,Dudin2012}. Measuring the occupation number of the optical mode performs a projective measurement of the qubit state.

The experimental sequence is illustrated in Fig.~1(c), see also  Refs.~\cite{Pritchard2011,Maxwell2013,Busche2017,Mohl2020a,Jiao2020}. The state $\vert {\sf 0}\rangle$ is initialised using a probe (red) and coupling (blue) laser with Rabi frequencies $\Omega_{\rm p}$ and $\Omega_{\rm c}$ to drive the two-photon transition transition  
$\vert{\rm g}\rangle\rightarrow\vert{\rm e}\rangle\rightarrow\vert{\rm r}\rangle$. 
Subsequently, we apply the microwave field, yellow in Fig.1(c), with Rabi frequency $\Omega_\upmu$ to drive the qubit transition. Finally, the coupling laser is turned back on to perform the read-out of state $\vert {\sf 0}\rangle$. The atomic ensemble of $^{87}{\rm Rb}$ atoms is laser cooled and transferred to an optical tweezer trap with wavelength 862~nm, beam waist of $w_0 =$ \SI{5}{\micro \meter} and trap depth $\sim 0.5~{\rm mK}$. The ensemble is cooled to \SI{50}{\micro \kelvin} and optically pumped into the state $\vert{\rm g}\rangle=\ket{5S_{1/2}, F = 2, m_F = 2}$. We load a few thousand atoms prior to any loss due to photon scattering events in order to achieve the requisite OD $\approx 4$ for photon storage and retrieval. Details on the preparation and optical response of our dipole traps can be found in references \cite{Busche2016,Bettles2018_collective}.
%
%
%
\begin{figure*}
    \centering
    \includegraphics[page=1]{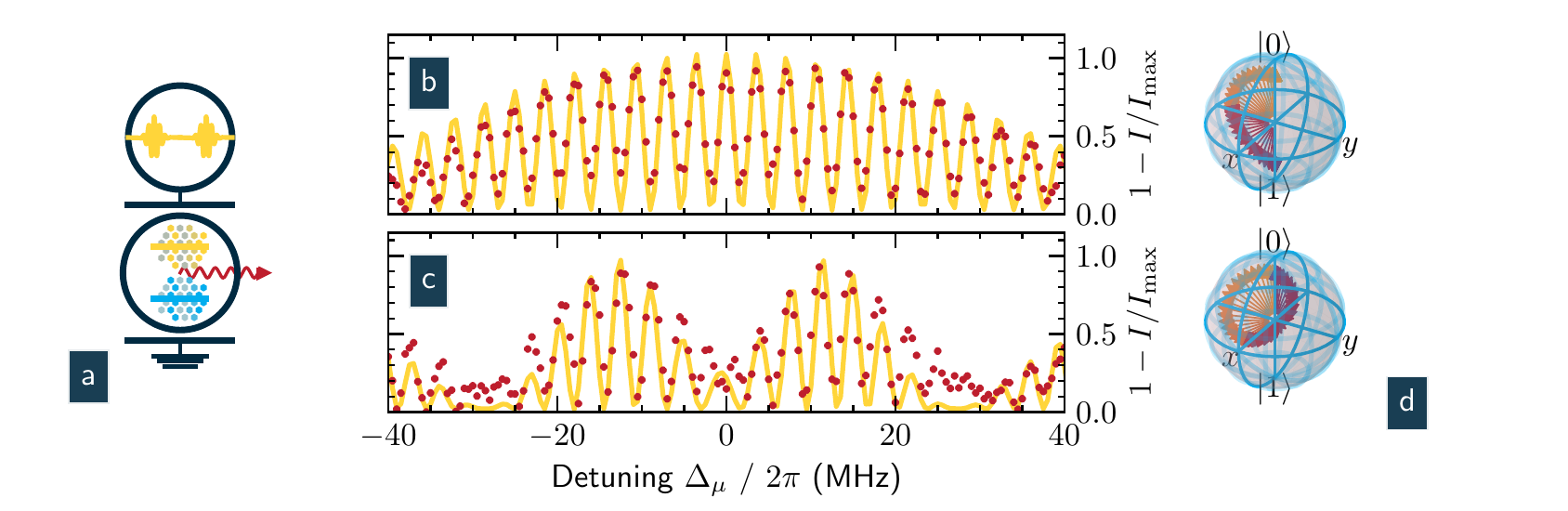}
    \caption{{\bf Microwave Manipulation of Rydberg Qubits: Ramsey Interferometry}. {\bf a}: The qubit is driven by two microwave pulses of duration 30-50 ns separated by a time $t_\mathrm{int}=$ \SI{250}{\nano \second}. The polariton retrieval protocol converts population in $\ket{0}$ to photons (red arrow), which are counted. {\bf b}: Normalised photon counts $I/I_\mathrm{max}$ (red circles) as a function of the $\pi/2$ pulse detuning, $\Delta_\upmu$, for the case of $\Omega_\upmu t_{\upmu}= \pi / 2$ at $\Delta_\mu=0$. A Monte Carlo simulation of the data is overlaid (yellow line). {\bf c}: The same as {\bf b} except  that $\Omega_\upmu t_{\upmu} =\sqrt{2}\pi / 2$, such that at $ \vert \Delta_\upmu\vert=\Omega_\upmu =2\pi$(\SI{14}{\mega\hertz}), we obtain a Hadamard gate. {\bf d} The evolution on the Bloch sphere for a resonant Ramsey interferometry, and double Hadamard operations.}
    \label{fig:RamseyFringes}
\end{figure*}
The circularly polarized probe beam, generated by an external cavity diode laser (ECDL),  drives the $\vert{\rm g}\rangle=\ket{5S_{1/2}, F = 2, m_F = 2} \rightarrow \vert{\rm e}\rangle = \ket{5P_{3/2}, F' = 3, m'_F = 3}$ transition on resonance. This light co-propagates with the dipole trap and is focused to a \SI{1}{\micro \meter} beam waist at the centre of the atomic ensemble. Probe pulses have a mean photon number of $\sim 0.25$ photons. This preserves optical depth and thus allows multiple experiments to be performed on the same ensemble. The coupling light, resonant with the $\vert{\rm e}\rangle\rightarrow\vert{\rm r}\rangle=\ket{n S_{1/2}}$ transition, is produced by a frequency-doubled diode laser system. The coupling beam is focused to $w_0 =$ \SI{30}{\micro \meter}, and counter propagates with the probe. This coupling light is offset locked to a temperature stabilised optical cavity via electronic side-bands \cite{Thorpe2008, Legaie2018}, and can be tuned to address Rydberg states with principal quantum numbers
$n = 30-95$.  The blockade mechanism at high $n$ suppresses multiple Rydberg excitations such that the retrieved light is observed to have $g^{(2)}(t=0) = 0.42 \pm 0.02$ for initialisation and read-out of the  $\ket{\rm r} = \ket{60S_{1/2}}$. Single excitation purity can be enhanced by using higher lying Rydberg states to $g^{(2)}(t=0) \sim 0.15$ \cite{Busche2017}.   The efficiency of writing a polariton and retrieving a photon is between 0.5\% and 1\% for $n=60$. This limitation is imposed by motional dephasing, blockade, and finite ensemble optical depth.


Single-qubit rotations are driven by coupling the $\vert{\rm r}\rangle= \vert{nS_{1/2}}\rangle$ and $\vert {\rm r}'\rangle = \vert{n'P_{3/2}}\rangle$ Rydberg states using a 16~mm \textit{in-vacuo} quarter-wave microwave antenna. The microwave source has a range of of 0-\SI{40}{\giga \hertz}, driving single-qubit rotations for Rydberg qubits with $n>46$, with $80\%/20\%$ switching time of \SI{10}{\nano\second}. Further experimental details can be found in references \cite{Busche2016}. All subsequent data in this paper correspond to $\ket{\rm r} = \ket{60S_{1/2}}$ and $\ket{{\rm r}'} = \ket{59P_{3/2}}$.

Figure \ref{fig:RamseyFringes} demonstrates coherent manipulation of a collective qubit. We observe quantum interference through Ramsey inteferometry using $\pi/2$ pulses and Hadamard gates.
The microwave pulse sequence is shown in Fig.~\ref{fig:RamseyFringes}{\bf a} and Fig. \ref{fig:schematicAndPulseSequences}{\bf c} (bottom row). 
Two microwave pulses separated by $t_\mathrm{int}$ = \SI{250}{\nano \second} perform single-qubit rotations in the $\ket{\sf 0}$ and $\ket{\sf 1}$ basis. The retrieved photon counts $I$, normalised to the maximum retrieved counts $I_\mathrm{max}$, as a function of the microwave detuning, $\Delta_{\upmu}$, for two values of the microwave pulse duration, $t_{\upmu}$, are shown in Fig.~\ref{fig:RamseyFringes}{\bf b} and {\bf c}.
In Fig.~\ref{fig:RamseyFringes}{\bf b} the power, $P$, and duration, $t_{\upmu}$, of each microwave pulse are chosen to give $\Omega_\upmu t_{\upmu} =\pi / 2$. In this case 
the sequence of two $\pi/2$ rotations about the $x$ axis in the Bloch sphere, see Fig.~\ref{fig:RamseyFringes}{\bf d}(top), separated by a rotation about $z$ (free evolution) results in familiar Ramsey fringes. In Fig.~\ref{fig:RamseyFringes}{\bf c}  the pulse duration is increased to give $\Omega_\upmu t_{\upmu}=\sqrt{2}\pi / 2$. The special case, $ \Delta_\upmu =\Omega_\upmu$, drives a Hadamard rotation ($\pi$ rotation about a Bloch vector $45^\circ$ from the $z$ axis), see Fig.~\ref{fig:RamseyFringes}{\bf d} (bottom). Consequently, the maximum fringe visibility in Fig.~\ref{fig:RamseyFringes}{\bf c} is observed at $\vert \Delta_\upmu\vert =\Omega_\upmu=2\pi(12~{\rm MHz})$. 
The theoretical fit in Fig.~\ref{fig:RamseyFringes}{\bf b} and {\bf c} (yellow) is calculated by solving the two-level master equation for experimental parameters. We assume that the Rydberg state lifetime is long compared to the experimental timescale and that motional dephasing can be neglected due to post-selection and normalisation.


\begin{figure*}
    \centering
    \includegraphics[page=1]{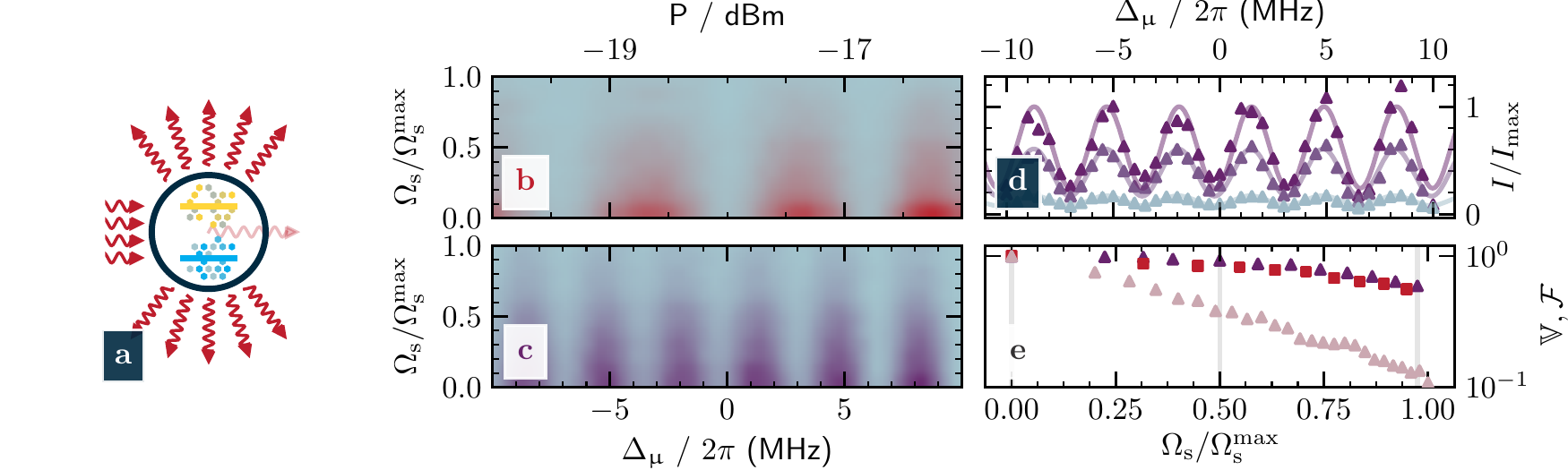}
    \caption{{\bf Robustness of the collectively-encoded Rydberg qubit to a non-Hermitian perturbation}: {\bf a}: A scattering field with amplitude $\Omega_{\rm s}$ is applied with wave vector $\mathbf{k}_{\rm s}$ along the photon read-out axis.  \textbf{b}: Rabi oscillation data: Heat-map of normalised photon counts as a function of microwave drive power ($P$) and scattering field amplitude $\Omega_{s}$, up to $\Omega_{\rm s}^{\rm max} / 2\pi =   $\SI{1.6}{\mega\hertz}. Red and blue indicate high and low photon counts, respectively. {\bf c}: Ramsey fringes data: Heat-map of normalised photon counts vs. microwave detuning for increasing $\Omega_s$. Purple and blue indicate high and low photon counts, respectively. {\bf d}: Selected Ramsay fringe data at $\Omega_{\rm s}/ 2\pi =0$, 1 and \SI{2}{\mega\hertz} (see also vertical grey bars in {\bf e}). {\bf e}: The visibility (${\mathbb V}$) of Rabi oscillations (red squares) and Ramsey fringes (purple triangles) as a function of $\Omega_{\rm s}$. The amplitude of polariton retrieval (read-out fidelity) (pink diamonds), is degraded significantly faster that the visibilities. Data in \textbf{b}, \textbf{c}, \textbf{d} are normalised to account for storage/retrieval efficiency $I_\mathrm{max}$. ${\mathbb V}$ and ${\cal F}$ in \textbf{e} are normalised to ${\mathbb V}_0$ and $\mathcal{F}_0$, the visibility and fidelity at $\Omega_{\rm s}$ = 0.
    }
    \label{fig:robustnessOfPhotons}
\end{figure*}

Next, to test the robustness of our collective encoding scheme, we apply a perturbation, ${\cal H}_{\rm p}$ during resonant Rabi oscillations or Ramsey interferometry,   see Fig. \ref{fig:schematicAndPulseSequences}{\bf c}.
First, we explore a non-Hermitian perturbation,
irreversibly removing atoms from the polariton by applying a scattering field, with amplitude $\Omega_{\rm s}$ resonant with $\vert{\rm g}\rangle\rightarrow\vert{\rm e}\rangle$, directed along the photon emission axis, see Fig.~\ref{fig:robustnessOfPhotons}{\bf a}. Figure~\ref{fig:robustnessOfPhotons}{\bf b} and {\bf c}
illustrate the loss of visibility of Rabi oscillations and Ramsey fringes as a function of $\Omega_s$. The visibility ${\mathbb V}$ is defined as the difference between the peak and minimum signals normalised by their sum.  Figure~\ref{fig:robustnessOfPhotons}{\bf d} shows the Ramsey fringes for low, intermediate and high values of $\Omega_s$. Figure~\ref{fig:robustnessOfPhotons}{\bf e} shows the visibility of the Rabi oscillation and Ramsey fringes plus the normalised amplitude of the retrieved mode (pink triangles), labelled $\mathcal{F}$ for fidelity. In the Supplemental Material \cite{supplemental} the photon retrieval fidelity $\mathcal{F}$ is calculated from the Lindblad equation for a single stored Rydberg polariton driven by the scattering field for duration $t$,
\begin{equation}
    \mathcal{F}= \exp\left(-4\frac{\Omega_{\rm s}^2 }{ \gamma_{\rm eg}}t\right)~.
    \label{eqn:fidelity}
\end{equation}
Here, $\gamma_{\rm eg}$ is the lifetime of the state $\ket{{\rm e}}$. The Rydberg state lifetimes $\tau_r, \tau_{r'}$ are assumed greater than the experimental timescale. Motional dephasing is omitted as experimental $\mathcal{F}$ are normalised to $\mathcal{F}(\Omega_s = 0)$. The data are in good agreement with the model, see Fig.~S1. The exponential dependence apparent in Fig.~\ref{fig:robustnessOfPhotons}{\bf e} arises due to the averaging over many runs using the same ensemble.

The main result of Fig.~\ref{fig:robustnessOfPhotons}{\bf e} is that reducing the polariton retrieval amplitude (thus $\mathcal{F}$) by an order of magnitude only reduces the qubit coherence, characterised by the Ramsey fringe visibility, by factor of two.  Hence we lose atoms from the polariton mode without significant degradation of the qubit coherence. In contrast to single-atom qubits where all the information is lost if a single atom is lost, our collective qubit is robust to atom loss.


%
%

\begin{figure}
    \centering
    \includegraphics[page=1]{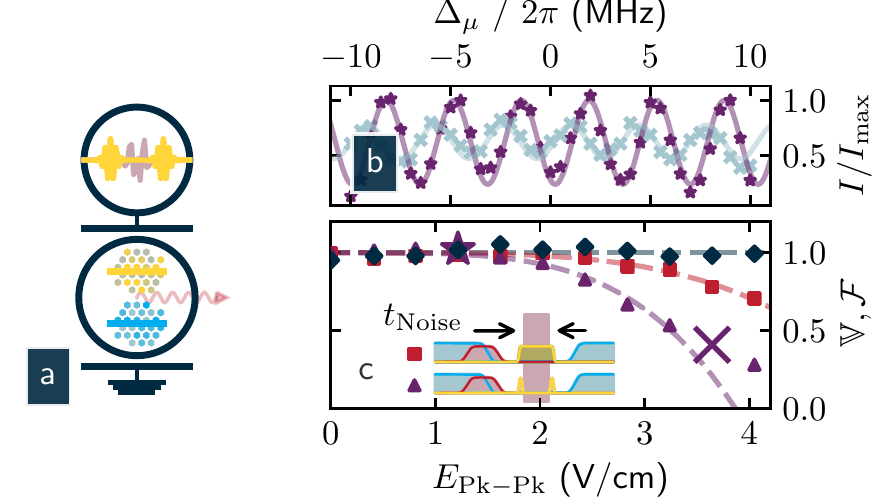}
    \caption{{\bf Noise-induced dephasing.} \textbf{a}:  A pulsed $E$-field with a 
    guassian amplitude noise  is applied during Ramsey and Rabi pulse sequences. The final state of the qubit is measured. \textbf{b}: Ramsey fringes corresponding to the central region of Fig.~\ref{fig:RamseyFringes} for low noise
    (${E}_\mathrm{Pk-Pk} = 1.2~$ V/cm, purple stars) and high noise (${E}_\mathrm{Pk-Pk} = 3.6~$ V/cm, blue crosses). Solid lines are sinusoidal fits. \textbf{c, main}: Rabi oscillation (red squares) and Ramsey fringe (purple triangles) visibility as a function of noise amplitude. Applied noise does not affect Fidelity (black diamonds). Both Rabi oscillation and Ramsey fringe visibility is proportional to $E_0^4$ (red, {\color{d_Purple}purple} dashed line), as predicted by the model. Ramsey visibility diverges from the quartic model at large ${E}_\mathrm{Pk-Pk}$ due to complex stark shifts. The star and the cross show datasets detailed in \textbf{b}.
    {\bf c, inset}: The duration of the pulse, $t_\mathrm{Noise}$  is equal to total duration of Rabi or Ramsey sequence, see Fig \ref{fig:schematicAndPulseSequences} {\bf c}. $\mathbb{ V}$ and ${\cal F}$ in \textbf{c} are normalised to ${\mathbb V}_0$ and $\mathcal{F}_0$, the visibility and fidelity at ${E}_\mathrm{Pk-Pk}$ = 0 V/cm. }
    \label{fig:noise}
\end{figure}

Finally, we test the robustness of our collective encoding scheme against decoherence induced by environmental noise. 
Using a Tektronic AFG3252 arbitrary wave form generator with bandwidth $240$~MHz, we apply an electrical noise pulse with peak-to-peak amplitude $E_\mathrm{Pk-Pk} = 2E_0$ to \textit{in-vacuo} electrodes, see Fig.~\ref{fig:noise}{\bf a}. The noise pulse perturbs Rydberg energy levels via the Stark effect and hence affects the global phase evolution of the collective states $\ket{0}$ and $\ket{1}$. This induces a $T2$ type decay similar to the thermally-induced decoherence in solid-state qubit systems~\cite{Chenu2017}. 
The applied noise field can be modelled as a static field term, giving rise to a quadratic Stark shift, and a fluctuating field, causing a dephasing
\begin{equation}
\mathcal{H'} = \frac{1}{2}
\begin{pmatrix}
\alpha_{\rm r'}&0\\
0&\alpha_{\rm r}
\end{pmatrix}
\frac{E_0^2}{3}+\frac{1}{2}
\begin{pmatrix}
\alpha_{\rm r'}&0\\
0&\alpha_{\rm r}
\end{pmatrix}\xi(t),\label{eqn:randomPerturbation}
\end{equation}
where $\alpha_j$ is the polarisability of $\vert j\rangle$, and $\xi (t)$ represents the fluctuations of $E^2$ about the average value $E^2_0/3$. Under the assumption of fast noise correlation decay, $\gamma_\mathrm{corr}=1/\tau_\mathrm{corr}$ larger than other decay mechanisms, the dephasing rate,
\begin{equation}
    \gamma_\mathrm{dep} =   \frac{4}{45}\cdot \frac{(\alpha_{\rm r'}-\alpha_{\rm r})^2\, E_0^4}{\hbar^2}\cdot\tau_\mathrm{corr}~,
    \label{eqn:electronicDephasing}
\end{equation}
where $\tau_\mathrm{corr}$ is the noise correlation time, see Supplemental Material \cite{supplemental}. The Stark shift is observed experimentally and is apparent in Fig.~\ref{fig:noise}{\bf b}.
The predicted $ E_0^4$-scaling is fit to the data in Fig.~\ref{fig:noise}{\bf c}. In the Supplemental Material, see Fig.~S2, we show that this quartic power law is a good fit. The data diverge from the quartic model at higher $E_0$, where the simplifying assumption of a quadratic Stark shift breaks down. 

%
%

In summary, we propose and demonstrate a novel collective encoding scheme for qubits based on Rydberg polaritons. We demonstrate fast coherent control using microwave fields. We find Rydberg qubits to have excellent coherence properties allowing for the implementation of fast Rabi oscillations, Ramsey interferometry and Hadamard gates. By performing Ramsey interferometry we demonstrate the robustness of a collectively encoded Rydberg qubit to depletion of atoms and to electric field noise. Rydberg qubits retain their quantum information even as the polariton suffers a partial loss of spatial phase coherence. We  demonstrate that Rydberg qubit dephasing due to electrical nosie depends quartically on the noise amplitude in agreement with theoretical predictions. Enhanced resilience to electrical noise might be obtained by utilising `magic' Rydberg states, where the polarisabilities of the Rydberg states are matched. Further work will focus on multiple qubits \cite{Busche2017}, qutrits, and phase gate proposals implemented using collective qubits \cite{Paredes-Barato2014, Khazali2019}.



\begin{acknowledgments}
    This work was supported by Engineering and Physical Sciences Research Council (EPSRC) Grants EP/M014398/1, EP/R002061/1 and EP/S015973/1. I.L. acknowledges support from the Deutsche Forschungsgemeinschaft through SPP 1929 (GiRyd) under Project No. 428276754, and “Wissenschaftler-Rückkehrprogramm GSO/CZS” of the Carl-Zeiss-Stiftung and the German Scholars Organization e.V.
\end{acknowledgments}

\bibliographystyle{apsrev4-1}

\end{document}


\maketitle

\section{Qubit Response to Photon Scattering}

In this section we provide the theory and additional results in support of Fig.~3 in the paper. When the collectively encoded qubit is resonantly driven on the $\ket{g} \leftrightarrow \ket{e}$ transition by a laser pulse (referred to as a scattering pulse in the main text) with Rabi frequency $\Omega_s$, the density matrix $\rho$ of each atom undergoes dissipative dynamics set by the master equation
\begin{align}
\partial_t{\rho} &= -{\rm i}[\mathcal{H}, \rho] + \gamma \mathcal{L}\rho\mathcal{L}^\dag - \frac{\gamma}{2}\left\{  \mathcal{L}^\dag\mathcal{L},\rho\right\},\\
\mathcal{H} & = \frac{\Omega_s}{2} \left(\ket{g}\bra{e}+\ket{e}\bra{g}\right),\\
\mathcal{L} & = \ket{g}\bra{e}.
\end{align}
The laser pulse is turned on for a given time $t$. It is then switched off ($\Omega = 0$) at which point all $\ket{e}$ population decays to $\ket{g}$.

In the following derivation we will calculate the temporal evolution of a Rydberg polariton shared amongst three atoms, and we will then generalise our results to larger ensembles. To prepare the Rydberg polariton we first initialise the atoms in the spin wave $\ket{e} = \frac{1}{\sqrt{3}}(\ket{gge}+\ket{geg}+\ket{egg}$ which is then adiabatically transferred to the polariton state $\ket{r} = \frac{1}{\sqrt{3}}(\ket{ggr}+\ket{grg}+\ket{rgg})$.

The density matrix of the initial state $\ket{i}$ is

\begin{equation}
\ket{i}\bra{i} = \frac{1}{3} \left(\ket{ggr}\bra{ggr} +\ket{ggr}\bra{grg} +\ket{ggr}\bra{rgg} + ... \right),
\end{equation}

where $\ket{ggr}\bra{ggr} = \ket{g}\bra{g}\otimes\ket{g}\bra{g}\otimes\ket{r}\bra{r}$. The figure of merit quantifying the robustness of the Rydberg polariton is $\mathcal{F} = \braket{i|\rho_\mathrm{f}|i}$, which defines the readout fidelity. Here $\rho_\mathrm{f}$ is the density matrix after evolution during the laser pulse. In the case of no dissipative dynamics the fidelity is $1$ for all times (this assumes that no other dissipative effects are present).

To calculate $\mathcal{F}$, we notice that the element $\ket{r}\bra{r}$ of the density matrix does not evolve, since dissipation acts in a different subspace. On the other hand, $\ket{e}\bra{e}$ does evolve in time, but decays back to $\ket{g}\bra{g}$ after the laser is switched off. This process is much faster than the experimental timescale. Thus the only element of the above density matrix with non trivial dynamics is $\ket{g}\bra{r}$.

\begin{align}
\partial_t{\ket{g} \bra{r}} &= -{\rm i} \Omega \ket{e}\bra{r},\\
\partial_t{\ket{e} \bra{r}} &= -{\rm i} \Omega \ket{g}\bra{r} - \frac{\gamma}{2} \ket{e}\bra{r}.
\end{align}

Solving these coupled differential equations with the initial conditions $\ket{g}\bra{r}_\mathrm{t=0}= \ket{g}\bra{r}$ and $\ket{e}\bra{r}_\mathrm{t=0}= 0$ gives

\begin{equation}
\ket{g}\bra{r}_\mathrm{t} = {\rm e}^{-\frac{\gamma}{4}t }\left(\cosh (\omega t)+\frac{\gamma}{4\omega}\sinh (\omega t)\right) \ket{g}\bra{r} - {\rm i} e^{-\gamma t}\frac{\Omega}{\omega}\sinh(\omega t)\ket{e}\bra{r},
\end{equation}

where $\omega = \frac{1}{4}\sqrt{\gamma^2- 16 \Omega^2}$. When switching of the laser all $\ket{e}$ population decays to $\ket{g}$ leaving only the $\ket{g}\bra{r}$ coherence, such that the application of the scattering pulse effects the following transformation:

\begin{align}
\ket{g}\bra{r} &\rightarrow e^{-\frac{\gamma}{4} t}\left(\cosh (\omega t)+\frac{\gamma}{4\omega}\sinh (\omega t)\right) \ket{g}\bra{r}= \alpha(t) \ket{g}\bra{r}
\label{eqn:coherenceTimeEvolution}
\end{align}

So the three-atom density matrix undergoes the transformation

\begin{align}
\ket{i}\bra{i} &\rightarrow \frac{1}{3}\left(\ket{ggr}\bra{ggr}+\ket{g}\bra{g}\otimes\alpha(t)\ket{g}\bra{r}\otimes\alpha^*(t)\ket{r}\bra{g} + .... \right) \\
& = \frac{1}{3}\left(\ket{ggr}\bra{ggr}+|\alpha^2(t)|\ket{ggr}\bra{grg}+..\right).
\end{align}

The structure of this final density matrix $\rho_\mathrm{f}$ becomes more apparent in matrix form:

\begin{equation}
\rho_\mathrm{f} = 
\frac{1}{3}\begin{pmatrix}
1 & |\alpha(t)|^2 & |\alpha(t)|^2\\
|\alpha(t)|^2 & 1 & |\alpha(t)|^2\\
|\alpha(t)|^2 & |\alpha(t)|^2 & 1
\end{pmatrix}.
\end{equation}

This result can then be generalised to N particles, where

\begin{equation}
\rho_\mathrm{f} = \frac{1}{N}|\alpha(t)|^2\begin{pmatrix}
1& . & . &1\\
. & 1 & . & .\\
. &.&1&.\\
1&.&.&1
\end{pmatrix}+\frac{1}{N}\left(1- |\alpha(t)|^2\right)
\begin{pmatrix}
1& . & . &0\\
. & 1 & . & .\\
. &.&1&.\\
0&.&.&1
\end{pmatrix}.
\label{eqn:coherenceFidelity}
\end{equation}
Evaluating the fidelity then yields
\begin{equation}
\mathcal{F} = \braket{i|\rho_\mathrm{f}|i} =   \frac{1}{N} + \frac{N-1}{N} |\alpha(t)|^2.
\end{equation}
Which for $\Omega \gg \gamma$ and $N\gg 1$ reduces to
\begin{equation}
    \mathcal{F} = \exp\left[-4 \frac{\Omega_s^2}{\gamma_s} t\right].
    \label{fidelity}
\end{equation}

Experimental results in Fig.~S1 support this theory. A qualitative match to the above prediction of an decay of the scattering field as a function of $\Omega_s$ is observed.

\begin{figure}
    \centering
    \includegraphics{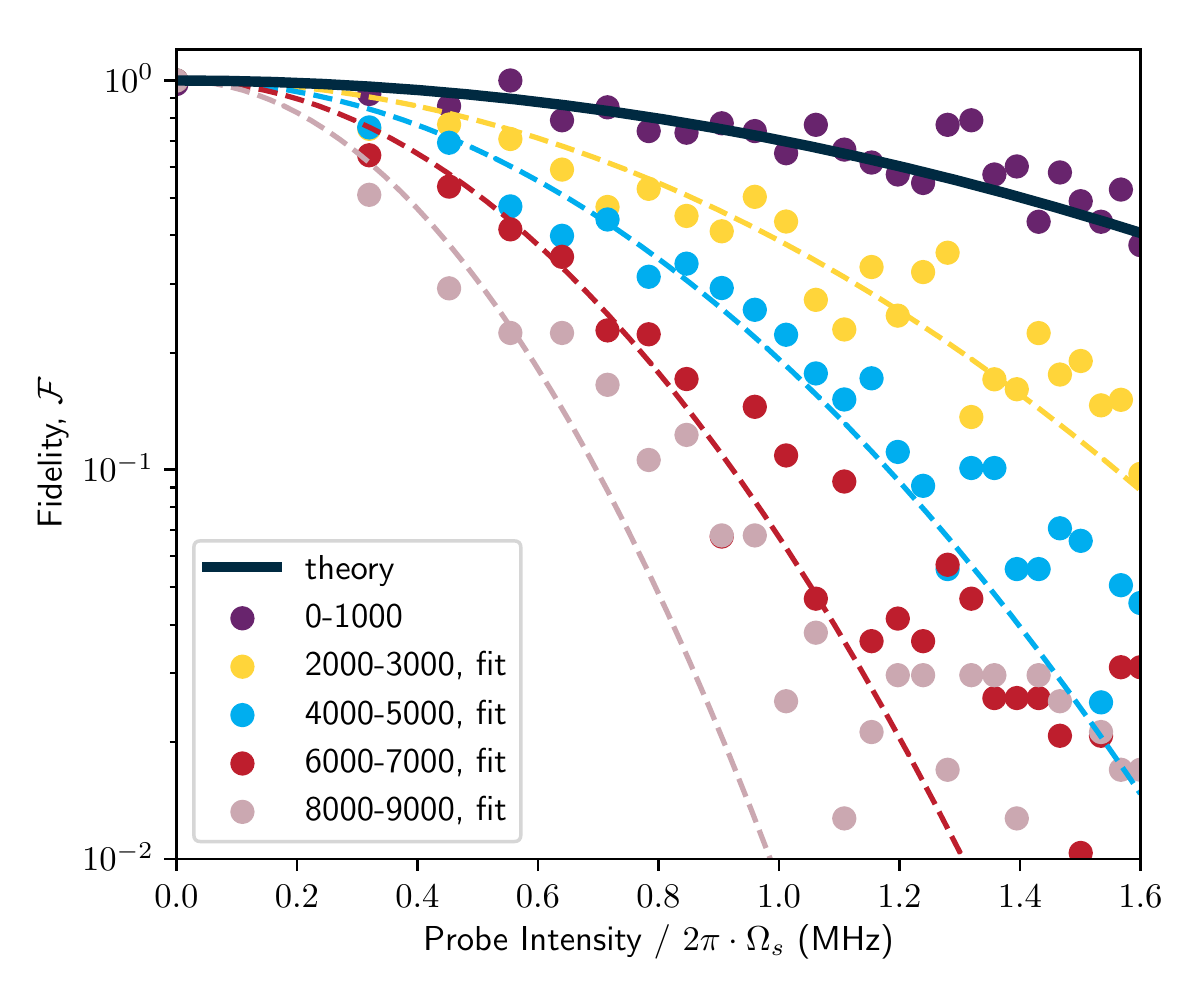}
    \caption{As the intensity of the scattering field is increased, the magnitude of the retrieval decreases due to a combination of loss of OD and fidelity reduction. Each ensemble is recycled for ten thousand experiments and the scattering field causes a progressive loss of optical depth which affects the magnitude of the retrieval. By grouping the data by shot number (1 being the first experiment performed upon an ensemble and 10,000 being the last, see legend inset) we are able to show that at low shot numbers, where OD is high, our observed retrieval converges upon the theory outlined in equation \ref{fidelity}. }
    \label{fig:imagesprobeIntensityVsShots}
\end{figure}

\section{Qubit Response to Electrical Noise}

In this section we provide theory and additional results in support of Fig.~4 in the paper. 
The coupling of the atomic Rydberg states to the externally applied electric field is modelled as a quadratic Stark perturbation described by the Hamiltonian
\begin{equation}
\mathcal{H} = \frac{1}{2}\begin{pmatrix}
\alpha_{\rm r'}& 0 \\
0 & \alpha_{\rm r}
\end{pmatrix}E^2(t).
 \label{eqn:perturbativeHamiltonian}
\end{equation}
Here $\alpha_{\rm r,r'}$ are the polarisabilities of the Rydberg $\ket{r}$ and $\ket{r'}$ states of the collective qubit. The electric field strength $E(t)$ is assumed to take the form of random variable over the interval $[-E_0, E_0]$.

Application of this perturbation introduces a relative, time dependent energy shift between the Rydberg states $\ket{r}$ and $\ket{r'}$.  It also leads to decoherence of quantum superpositions of states in $\ket{r}$ and $\ket{r'}$. To discriminate between the two effects, we can decompose the squared electric field term in equation \ref{eqn:perturbativeHamiltonian} as
\begin{equation}
E^2(t) = \braket{E^2(t)} + \xi(t).
\end{equation}
Here $\braket{E^2(t)}$ is the time-averaged value of $E^2(t)$
\begin{equation}
\braket{E^2(t)}=\frac{1}{2E_0}\int_{-E_0}^{E_0} E^2(t)\, dE(t) = \frac{E_0^2}{3}.
\end{equation}
$\xi(t) = E^2(t) -  \braket{E^2(t)}$ represents time dependent fluctuations of $E$ about this mean value.

The Hamiltonian of equation \ref{eqn:perturbativeHamiltonian} can now be rewritten as a sum of a constant part (causing the quadratic Stark shift) and a time-dependent perturbation which is a fluctuating term with mean value zero:
\begin{equation}
\mathcal{H} = \frac{1}{2}\begin{pmatrix}
\alpha_{\rm r'} & 0 \\
0 & \alpha_{\rm r}
\end{pmatrix}\frac{E^2_0}{3}
+ \frac{1}{2}\begin{pmatrix}
\alpha_{\rm r'} & 0 \\
0 & \alpha_{\rm r}
\end{pmatrix}\xi(t).
\end{equation}

The temporal correlations of $\xi(t)$ can be rewritten in terms of $E(t)$:
\begin{eqnarray*}
\braket{\xi(t)\xi(t')} &= &\braket{(E^2(t) - \braket{E^2(t)})(E^2(t') - \braket{E^2(t')})}\\
& =& \braket{E^2(t)E^2(t')}-2\braket{E^2(t)}\braket{E^2(t')}+\braket{E^2(t)}\braket{E^2(t')}\\
& = &\braket{E^2(t)E^2(t')}-\braket{E^2(t)}^2.
\end{eqnarray*}
Then, using 
\begin{equation}
\braket{E^2(t)E^2(t')}=\frac{1}{2E_0}\int_{-E_0}^{E_0} E^2(t)E^2(t') dE(t) = \frac{E_0^4}{5}
\end{equation}
we find that for equal times ($t=t^\prime$):
\begin{equation}
\braket{\xi(t)\xi(t)} = \frac{E_0^4}{5} - \frac{E_0^4}{9} = \frac{4}{45}E_0^4.
\end{equation}
For unequal time, we assume that the correlation function decays exponentially with a correlation time $\tau_\mathrm{corr}$:
\begin{equation}
\braket{\xi(t)\xi(t')}  \approx \braket{\xi(t)\xi(t)} \exp{\left[-\frac{|t-t'|}{\tau_\mathrm{corr}}\right]} = \frac{4}{45}E_0^4 \exp{\left[-\frac{|t-t'|}{\tau_\mathrm{corr}}\right]}.
\end{equation}

We assume that the noise correlation time is much shorter than the typical timescales governing the evolution of the collective qubit. In this case we can approximate the rapidly decaying correlations with a delta-function:
\begin{equation}
\braket{\xi(t)\xi(t')} \approx 2 \tau_\mathrm{corr} \frac{4}{45}E_0^4\, \delta(t-t'),
\end{equation}
Where the factor of $2 \tau_\mathrm{corr}$ ensures that the integral over the noise correlations is unchanged. The evolution of the density matrix of the system
\begin{eqnarray}
\rho=\left(
  \begin{array}{cc}
\rho_{\rm r'r'} & \rho_{\rm rr'}\\
\rho_{\rm r'r} & \rho_{\rm rr}\\
\end{array}
\right)
\end{eqnarray}
is then obtained via a Lindblad master equation of the form
\begin{equation}
\frac{\partial}{\partial t}\rho(t) = -\frac{{\rm i}}{\hbar}\left[\mathcal{H}, \rho(t)\right]+\mathcal{D}(\rho(t)).
\end{equation}
After Chenu et al \cite{Chenu}, the dissipator $\mathcal{D}(\rho(t))$ is given by
\begin{align}
\mathcal{D}(\rho(t))&= -2 \tau_\mathrm{corr} \frac{4}{45}E_0^4 \frac{1}{\hbar^2}
\left[\frac{1}{2}
\left(
  \begin{array}{cc}
\alpha_{\rm r'} & 0 \\
0 & \alpha_{\rm r}\\ 
\end{array}
\right), \left[\frac{1}{2}\left(
  \begin{array}{cc}
\alpha_{\rm r'} & 0 \\
0 & \alpha_{\rm r}\\ 
\end{array}
\right), \rho(t)\right]\right]\\
& = - \frac{1}{2} \tau_\mathrm{corr} \frac{4}{45}E_0^4 \frac{1}{\hbar^2}
\left(
  \begin{array}{cc}
0 & (\alpha_{\rm r}-\alpha_{\rm r})^2\rho_{\rm rr'}\\
(\alpha_{\rm r'}-\alpha_{\rm r})^2\rho_{\rm rr'} & 0\\
\end{array}
\right).
\end{align}
Coherences between the qubit states are expressed by the operator 
\begin{eqnarray}
\sigma_x=\left(
  \begin{array}{cc}
0 & 1\\
1 & 0\\
\end{array}
\right),
\end{eqnarray}
which evolves according to 
\begin{equation}
\frac{\partial}{\partial t}\sigma_x(t) = -\frac{\gamma_\mathrm{deph}}{2}\sigma_x(t),
\end{equation}
where
\begin{equation}
\gamma_\mathrm{deph} = \tau_\mathrm{corr} \frac{4}{45}\cdot \frac{(\alpha_{\rm r'}-\alpha_{\rm r})^2\, E_0^4}{\hbar^2}.
\end{equation}
is the dephasing rate.\\

In order to investigate the dependence of shift and dephasing on the amplitude of electrical noise applied to the collective qubit, the qubit was modelled as a two level system as with equation \ref{eqn:perturbativeHamiltonian}, where an energy offset is included to give a symmetric symmetric level shift $\Delta_\mathrm{int} = (\alpha_{\rm r'} - \alpha_{\rm r})E_0^2 / 3$ and dephasing rate $\gamma_\mathrm{dep}$.  

\begin{equation}
\dot\rho=-\frac{{\rm i}}{\hbar}
[\mathcal{H},\rho]
+ \gamma_\mathrm{dep}
\left[
L\rho L^\dagger -\frac{1}{2} \left\{
L^\dagger L, \rho
\right\}
\right]
\label{eqn:lindblad}
\end{equation}

where the Hamiltonian of the qubit was parameterised with both the interferometer shift, and the detuning of the microwave field coupling $S$ and $P$.

\begin{equation}
\mathcal{H} = \frac{1}{2}
\begin{pmatrix}
\Delta_{\rm int} +\delta_\mathrm{int} & 0 \\
0 & -\Delta_{\rm int}-\delta_\mathrm{int}
\end{pmatrix}
\end{equation}

Fringe visibilities were used to determine the degree of dephasing. Visibilities are calculated with simple sinusoidal fits of the form $A_\mathrm{fit} \sin (w_\mathrm{fit}*\Delta_{\rm int} + \phi_\mathrm{fit}) +O_\mathrm{fit}$, where $A_\mathrm{fit}$ is the fringe amplitude, $w_\mathrm{fit}$ the frequency, $\phi_\mathrm{fit}$ the phase offset and $O_\mathrm{fit}$ is the amplitude offset.

\begin{equation}
\text{Vis} = A_\mathrm{fit}/O_\mathrm{fit}.
\end{equation}

This simple calculation compensates systematic errors arising due to shot-to-shot fluctuations in storage/retrieval efficiency observed during operation of the experiment.

\begin{figure}
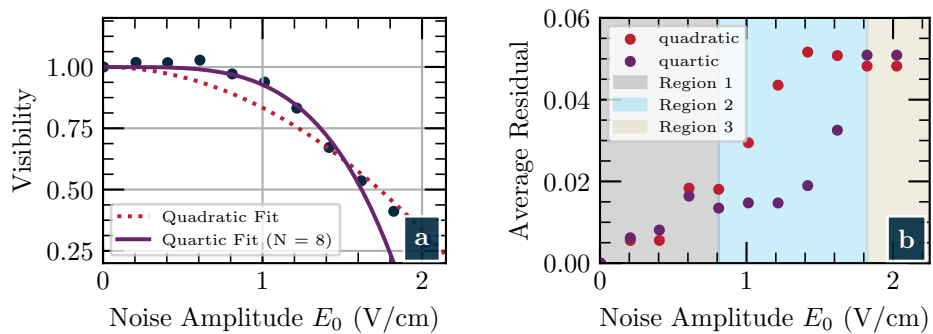

\input{./images/visibilityDataFitComparison.pgf}
\input{./images/dephasingQuadraticQuarticComparison.pgf}
\caption{Comparing Quadratic and Quartic models to fringe visibility. \textbf{a} Quadratic (red) and quartic (purple) fits to the normalised fringe visibility of the interferometer as a function of electric field.  The quadratic fit outperforms the quartic model when fitting the whole dataset. However the quartic model performs significantly better when fitting only the first N data points, where $N<8$ and the visibility remains above 50 percent. \textbf{b} Average residuals of quadratic (red) and quartic (purple) fits to the first N visibility data points. Three regions are highlighted in grey (R1, $E_\mathrm{0}<0.8$ V/cm), cyan (R2, $0.8$ V/cm $<E_\mathrm{0}<1.8$ V/cm), and gold (R3, $1.8$ V/cm$<E_\mathrm{0}$). In Region 2, a quartic dependence is a much better fit to the data, evidenced by smaller average residuals. In region 3, the visibility drops to 20 percent of the visibility at $E_0 = 0$. A departure from strong quartic scaling is observed as visibility approaches the limiting value of zero.}
\label{fig:visibilityQuartic}
\end{figure}

\begin{figure}
\includegraphics[width=\textwidth]{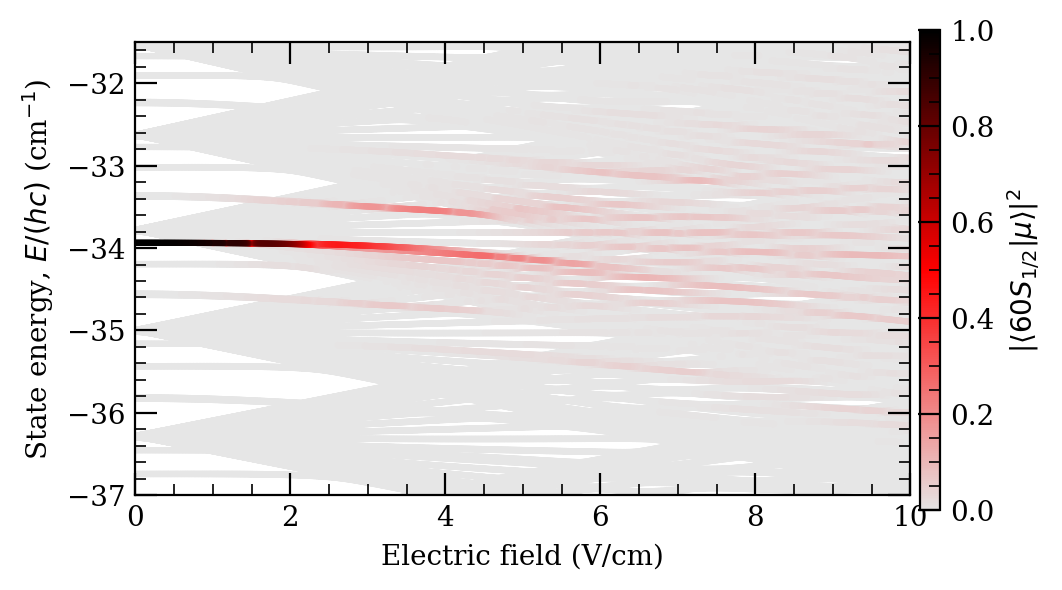}
\includegraphics[width=\textwidth]{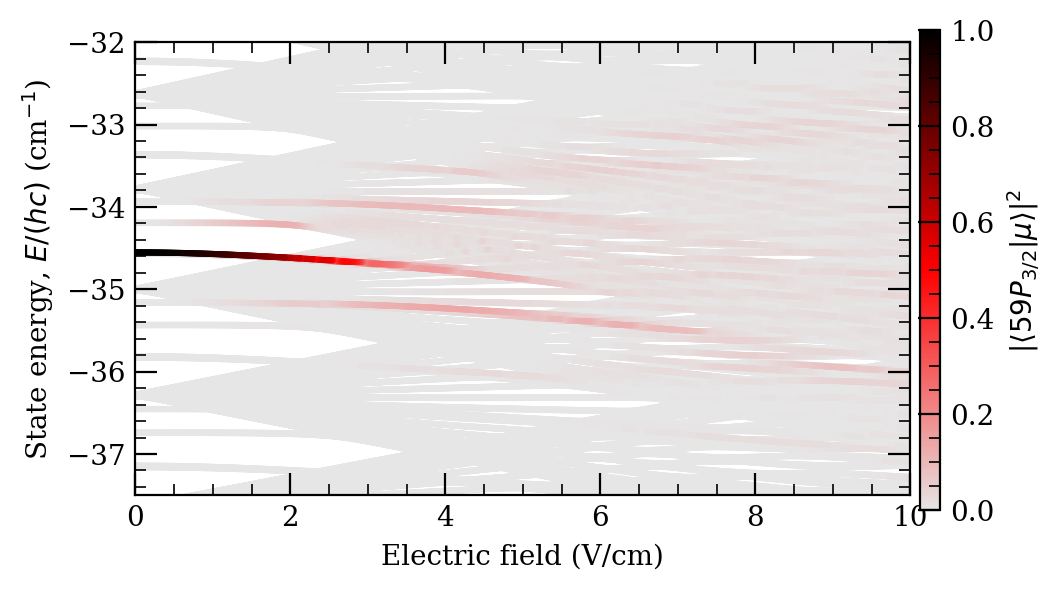}
\caption{Stark shifts of the $60S_{\frac{1}{2}}$,$59P_{\frac{3}{2}}$  energy level of Rubidium. The stark effect is only quadratic for small $E_0$. Divergence from a simple $E^2$ relationship begins at around around $E_0 = 2$ V/cm, and is clearly present at $E_0 = 4$ V/cm for both states. This effect can be seen in the divergence of the observed fringe visibility from the quartic model of figure \ref{fig:visibilityQuartic}.}
\label{fig:startShift60s0.5}
\end{figure}

In order to compare quartic and quadratic models, a series of fits were performed on datasets comprising of N data points spanning $E^{(0)} = 0$ V/cm to $E^{(n)}$, where  $E^{(n)}$ is the electric field at the $n$'th data point. The goodness of these fits is summarised in figure \ref{fig:visibilityQuartic}b. From these fits and residuals, it can be seen that the reduction in visibility is initially quartic, diverging from this relationship only at higher electric fields $E_0>2$ V/cm. Comparison of this data with the stark maps for Rubidium (see figure \ref{fig:startShift60s0.5}) show that the breakdown of the quartic model occurs due to the complex stark splitting with an onset of $E_0=4$V/cm. This complex stark splitting also causes a reduction in retrieval efficiency after exposure to strong electrical noise.

\bibliographystyle{naturemag}
\bibliography{references}